\documentstyle[12pt]{article}
\renewcommand{\thefootnote}{\fnsymbol{footnote}}
\begin{document}
\baselineskip 0.78cm
\raggedbottom
\renewcommand{\thefootnote}{\fnsymbol{footnote}}
\setcounter{footnote}{1}
\def \dbarsd  {$\bar d_{sea}^\downarrow (x)$}
\def \qq      {$q\bar q$}
\def \ss      {$s\bar s$}
\def \uu      {$(u\bar u+d\bar d)/\sqrt 2$}
\def \KK      {$K\overline K$}
\def \pieta   {$\pi\eta $}
\def \pipi    {$\pi\pi$}
\def \ff      {$f_1(1420)$}
\def \io      {$\eta(1440)$}
\def \fo      {$f_0(1520)$}
\def \f       {$f_2(1520)$}
\def \th      {$f_0(1710)$}
\def \S       {$f_0(980)$}
\def \a       {$a_0(980)$}
{\hfill HU-SEFT R 1994 - 03}
\bigskip

{\centerline {\bf HOW TO PARAMETRIZE AN S-WAVE RESONANCE}}
{\centerline {\bf AND HOW TO IDENTIFY TWO-HADRON COMPOSITES}}
\vskip 0.5cm
\centerline{NILS A. T\"ORNQVIST}
\centerline{\it
Research Institute for High Energy Physics (SEFT)}
\centerline{\it
PB 9, FIN-00014 University of Helsinki, Finland}

\begin{quote}
\small
{\baselineskip 0.8cm
The question of  how one can distinguish quark model states from
2-hadron states near an S-wave theshold is discussed,
and the usefulness of the running mass is  emphasized
as the meeting ground for experiment and theory and for
defining resonance parameters.}
\vskip .2cm

\end{quote}

{\bf 1. Introduction.}
A current problem of fundamental importance in hadron spectroscopy is:
How to distinguish composites formed of two hadrons from
normal quark model hadrons? There are now a handful of good experimental
candidates \cite{PDG}
which have great difficulties in finding a place whithin the
normal \qq\ model. Examples of such states are the \ff , \fo , \f ,
\th , and $\Lambda (1405)$
and a longstanding problem has been the question whether the
\S\ and the \a\ are \qq\ or \KK\ states. All these resonances
appear near an important S-wave threshold.

Recently Morgan and Pennington \cite{Mor} and Zou and Bugg \cite{Bugg}
have discussed the structure of the \S , noticing that one needs two nearby
poles to describe the \S .
Morgan and Pennington \cite{Mor}
made the important observation that this two pole
structure is what is expected from a normal \qq -meson near an S-wave
threshold (\KK ) in contrast to a \KK\ bound state, for which only one
pole is expected.

 This seems to provide a nice clear-cut method to almost model independently
distinguish hadron-hadron bound states from normal quark model states. I
shall here show that the question of whether one  has one pole
or two poles depends on the effective distance to left hand cuts,
or on the range of the binding forces.
I shall clarify the issues involved, first through
some remarks of general nature, and then by emphasizing the
usefulness of the concept of the running mass m(s) \cite{NAT}, which also
provides a good way to distinguish between a \qq\ and a hadron-hadron
state. The shape of
this running mass function allows one to in a simple single picture
discuss most of the problems involved, and at the same time to see the
origin of the one pole - two pole dichotomy.

We shall be concerned with S-wave thresholds, since for higher
orbital momenta  the centrifugal factor makes the important
 S-wave square root cusp disappear.
Almost all cases of practical interest are, in fact,
in S-wave channels.
The fact that a  \qq\ state sometimes requires
two poles brings to the fore another old problem of great importance for
experiment, which I  discuss at the end:
 How should one parametrize a resonance near threshold? - by the pole
positions or by Breit-Wigner (BW) parameters?
\bigskip

{\bf 2. Poles from \qq\ resonances.}
Let us start the discussion with normal \qq\ states. Since these have their
origin in the confined sector owing their binding to gluonic exchange, they
are CDD poles \cite{CDD}.  A school example of a CDD pole is the $K^0$ pole
in $\pi\pi\to\pi\pi$ due to the weak interaction $K^0\to\pi\pi$. Although
usually disregarded this makes,
in principle, the $\pi \pi$ phase shift jump by 180$^\circ$ at the
$K^0$ mass.
There is a formal similarity with this $K^0$ pole and a normal \qq\ state,
which couples by \qq\ pair creation to hadron-hadron channels. Both are
CDD poles and come from another sector of Hilbert space, than that spanned
by the decay channels. Only the magnitude of  the coupling to hadrons,
is orders of magnitudes stronger for \qq\ than for $K^0$.
If one could make the quark pair creation very small,
 like in a zero-width approximation or in a quenched approximation
of lattice QCD, the \qq\ poles would still remain as spikes in
hadron-hadron amplitudes, albeit shifted in mass from their normal
positions. In other words,
the quark loops (which together with the initial quarks make hadron
loops) will generally shift down the "bare \qq\ masses"
at the same time as the resonances
aquire finite widths. This is quite different from the case of
hadron-hadron bound states. There, if the coupling is decreased the
state disappears completely out of existence.

Let us imagine that we could tune the bare \qq\ mass
so that the true resonance
position passes the threshold. To be concrete, let us chose the
\KK\ threshold.
The I=1 channel is slightly
simpler than I=0, since we expect only one resonance, the $a_0(980)$,
whereas for I=0 one also has
energy dependent complex mixing \cite{NAT} between the \ss\ and \uu\
resonances. Near the threshold
the  inverse propagator has the  the general form:
\begin{eqnarray}
 P^{-1}(s)&=& m^2_0-s+\Pi (s)  \nonumber \\
&=& m^2(s) -s -
i\gamma'  \sqrt{s-4m_K^2}G(s) \Theta(s-4m_K^2)-im\Gamma_{\eta\pi}\ ,
\label{prop}
\end{eqnarray}
where $m^2_0-s$ comes from the bare \qq\ propagator
and $\Pi (s)$ is the "correction"
term due to \KK\ and other hadronic loops, whose imaginary part is
given in the  second expression.
The constant
$\gamma' $ measures the strength of the coupling to \KK\ and
$G(s)$ is a form factor which includes left hand cuts, and
if one wishes, a factor $2m_K /\sqrt s$ for relativistic phase space.
The simplest parametrization of $G(s)$ is by a pole such that
\begin{equation}
G(s)= [1+(s-4m_K^2)/\mu^2]^{-1} \ .
\label{ff}
\end{equation}
The constant  $\mu$ gives a cutoff, and is in  order of magnitude given by
the energy of the t-channel exchanges\footnote{If the factor $2m_K/\sqrt s$
from relativistic phase space is included in
eq.~(\ref{ff}) the slope of linear term in (\ref{runmass}) is increased by
$\gamma' /(m_K\pi)$. As in the discussion following eq.~(\ref{runmass})
this linear term can be absorbed
by renormalizing the coupling $\gamma'$.}.
Normally  for a \qq\ state this cutoff should be large.
The squared running mass
$m^2(s)$ in eq.~(\ref{prop}) is $m_0^2$ plus the mass shift function
Re$\Pi (s)$, which
is given by  Im$\Pi (s)$ through a dispersion relation

\begin{eqnarray}
m^2(s)&=& m^2_0+\frac 1 \pi {\cal P} \int\frac{{\rm Im}\Pi (s')}{s'-s}ds'
= m^2_0+\nonumber \\ &+&
 \gamma' [-\mu +(s-4m_K^2)/\mu +\sqrt{4m_K^2-s}\Theta(4m_K^2-s)]
+ {\cal O}(s-4m_K^2)^{\frac 3 2} .\label{runmass}
\end{eqnarray}
The negative slope of this running mass,
$\alpha=-dm^2/ds$, evaluated at the bound state pole is proportional to
the probability  ($Z$) to find the state as $K\overline K$, when the
\qq\ probability  is normalized to 1, i.e., $Z=\alpha/(1+\alpha)$.
Thus for a \qq\  state, shifted in mass
a little below an S-wave threshold, $Z$ can be
close to unity, since the slope diverges at the threshold.
Thus most of the time such a "unitarized remnant of a
\qq\ state" is found to be in a \KK\ state, while the
 \qq\ probability, $1-Z=1/(1+\alpha)$,  is small although
the state owes its existence to the \qq\ sector.
This  factor, $1/(1+\alpha )$, also appears in the \KK\ coupling constant
defined by the pole residue, $g_{K\bar K}^2/{4\pi} =\gamma' /(1+\alpha)$,
making the $g_{K\bar K}$  very sensitive to the exact pole position.

The function Re$\Pi (s)$ shifts the mass down
by $\approx \gamma' \mu$ and includes a
linear term, $s\gamma' /\mu $. This constant and
linear term can, however, be absorbed by the
corresponding terms of the inverse propagator eq.~(1)
(at least for small $\gamma' /\mu<1$, see discussion below),
by the redefinitions: $m^2_{BW}$ $=(m^2_0-\gamma' \mu)/(1-\gamma' /\mu)$
and  $\gamma =$$\gamma' /(1-\gamma' /\mu )$,
since the T matrix element,  $\gamma' G(s)P(s)$, actually
depends on only two independent
parameters, not on all three of eq.~(\ref{runmass}).
Thus all the essential features of an S-wave \qq\ propagator
near the threshold is described by the form Flatt\' e
used long time ago \cite{Flatte}:
\begin{equation}
m^2(s)=m^2_{BW}+\gamma \sqrt{4m_K^2-s}\Theta (4m_K^2-s)\ .\label{mBW}
\end{equation}
 For this simple function of a constant plus a square root cusp
the pole  traces a path in the complex plane
of the kaon c.m.~momentum $k_K$ shown in Fig.~1a
by the dotted lines. The filled black squares are special examples.
The straight lines are obtained assuming $\Gamma_{\eta\pi}=0$,
 while one gets the hyperbola for a small but finite $\Gamma_{\eta\pi}$.
The  \KK\ coupling  $\gamma $ is given in the figures
a rather realistic value of 0.4 GeV, which gives a typical
strong BW  width of about 200 MeV, if the resonance would be
200 MeV above threshold. For $\Gamma_{\eta\pi}=0$,
bound states lie on the positive  Im$k_K$ axis, and resonance poles lie
slightly below the positive real axis (Im$k_K=-\gamma /4$).
The shadow pole lies symmetrically reflected with
respect to the point $(0,-\gamma /4)$ and at the unfilled
squares for the examples. Normally these are far away from the physical
region.
imaginary axis for bound states).
But, when the \qq\ state is near the threshold (and in particular
if it is a virtual bound state) then the shadow pole creeps up very close to
the physical region, and influences the resonance shape, just as the true
pole does.

If one  allows a finite
$\Gamma_{\pi\eta }$ (which is slowly varying in the
narrow energy region of interest near the \KK\ threshold) then the
bound state pole is shifted to the left branch of the hyperbola in
Fig.~1a,
while the shadow pole is shifted to the right branch.
With this open \pieta\ channel
there are 4 sheets of the energy plane; two for each threshold.
Conventionally \cite{Bad}
these are numbered such that in the $k_K$ plane
(which is cut by the \pieta\ threshold
along the imaginary axis) each quadrant correspond to
sheets with Roman numbers shown in Fig.~1a.

The same poles discussed  in Fig.~1a
trace in the $s$-plane, Fig.~1b, much more complicated trajectories.
With increasing bare \qq\  mass
the physical \qq\ pole position (the solid line for
$\Gamma_{\pi\eta }=0)$
increases until one reaches the threshold. Then, it creeps through the cut
and becomes a virtual
bound state, for which the pole mass actually decreases, although the bare
mass increases! For such a virtual state
Im(pole)$=0$ and Re(pole)$<4m_K^2$, but
it is still seen  also  as a
BW resonance with a finite \KK\ width and phase shift passing
through 90$^\circ$ a little above threshold. Eventually, as the bare
mass increases
the pole aquires an imaginary part and Re(pole)  increases again, while the
resonance pole is along the lower branch of the parabola in Fig.~1b.

The shadow poles are also shown in Fig.~1b, by the dashed lines.
These are close to the physical region when the regular pole is near
threshold. The BW parameters
are also shown (when $\Gamma_{\pi\eta }=0$) by the filled circles
 and by the dotted half-parabola. As can be seen the pole parameters
[the zeroes of eq.~(\ref{prop})] and the
BW parameters [defined at the zeroes of the real part of eq.~(\ref{prop})]
can differ considerably.
With the value of $\gamma =0.4$ GeV,
as chosen in the figures, the BW mass is 50-100 MeV larger
than the pole mass, and
in particular a virtual bound state whose Im(pole)=0,
 can have a BW width of 100 MeV!

The remaining dotted curves in
Fig.~1b shows how the pole
in the $s$-plane is shifted when one adds a small
width, $\Gamma_{\pi\eta}$, as was done in Fig.~1a.
It is interesting to note that it is the shadow
of the bound state, and not the bound state itself,
which is continuously connected to the resonance pole
when one increases the mass of the state though the threshold region.
Similarily it is the bound state
(or the virtual bound state) pole which turns
into the shadow pole of the resonance. Thus, although below and far above
the threshold it is obvious which pole
should be chosen as the physical pole, a little above
the threshold both poles
are "equally physical", and it is not clear which of the poles is the
shadow pole! Near the threshold it is thus mandatory to give both poles
in order to have a complete description
of the pole structure.

In Fig.~2a $m^2(s)-m_{BW}^2$ of eq.~(\ref{mBW}) is shown
with the same parameters as in Figs.~1a-b,
when $\Gamma_{\eta\pi}=0$.
The crossing points with the line $s-m_{BW}^2$ gives the pole mass of the
state if it is below threshold,
while above threshold it simply gives  the BW mass of a resonance.
If the line crosses the shadow branch of the running mass
(dashed in Fig.~2a) one finds the shadow pole, and if there are
two crossings also the virtual bound state pole.
Thus one can find from the same plot the positions of both pole and
shadow pole, and for a virtual bound
state one can furthermore also find the
BW mass. For a resonance well above threshold the graph gives of course
only the BW mass. One sees easily  why two poles
appear near threshold, and why the shadow pole comes equally close to
the physical region as the true pole.

{\bf 3. Deuteronlike states.} In the discussion
above I have up til now assumed that the linear term in (\ref{runmass})
is small enough ($\gamma' /\mu<1$), so that it can be removed by the
redefinitions of $m_0$ and   $\gamma' $,
For a sufficiently steep form factor
this condition is violated, and the nature
of the solution changes dramatically.
The shorter the distance  ($\mu$) to the left hand singularities is, the
further moves the shadow pole  from the true pole
until for $\gamma' /\mu=1$ it is at $-\infty$.

This situation resembles, in fact, that of the deuteron (Fig.~2a),
or in general that of any 2-hadron bound state. There, the
bare \qq\ term of eq.~(\ref{prop}), $m^2_0-s$, is of course absent
and the inverse propagator is given by an analytic function $D(s)$,
which like $\Pi(s)$ above, is assumed to approach
a constant at $s\to\pm\infty$. When this
function developes a zero, $D(m_d^2)=0$,
below threshold one has a bound state
at $s=m_d^2$. The running mass can be written:
\begin{equation}
m_d^2(s)=s-D(s)/D'(m_d^2)=m_d^2-\frac{(s-m_d^2)^2}{\pi D'(m^2_d)}
{\cal P} \int \frac{{\rm Im}D(s')ds'}{(s'-m_d^2)^2(s'-s)}\ .\label{deutm}
\end{equation}
In Fig.~2a the parameters for the deuteron $m_d^2(s)$
are  obtained from the scattering length
$a=3.82m_\pi^{-1}$ and effective range parameter $r=1.2m_\pi^{-1}$
of the proton-neutron $^3S_1$ wave phase shifts above threshold.
With a left hand cut parametrized as  $G(s)$ of eq.~(\ref{ff})
the  parameter $\mu=2.68m_\pi$ and
the deuteron pole (more precisely, $(4m_N^2-m_d^2)^{\frac 1 2}$)
are then fixed by $a$ and $r$ through $2[1\pm (1-2r/a)^{\frac 1 2}]/r $.
Up to terms of ${\cal O}(s-4m_N^2)^{\frac 3 2}$ one has:
\begin{equation}
m_d^2(s)=s+ \frac{g_{dNN}^2}{4\pi}[-\frac{2}{a}+(s-4m_N^2)
(\frac r 4 +\frac 1{a\mu^2})+
\sqrt{4m_N^2-s}\Theta (4m_N^2-s)]\ , \label{deut} \\
\end{equation}
where $m_N$ is the nucleon mass, and  $g_{dNN}^2/(4\pi)$
is also fixed by $a$ and $r$ through  the condition
$dm^2(s)/ds=0$ at the deuteron pole.

Notice that {\it for a deuteronlike state
 the slope of the running mass above threshold must be
larger than unity}, as can be seen
from eq.~(\ref{deutm}). This  gives the condition,
$r+4/(a\mu^2)>0$, or in practice that the effective range parameter $r$
is positive \cite{Wein}, in contrast to the case of a \qq\ state where
$r=-8/\gamma$ is negative.
This also implies
that there is no crossing with $s$ and the shadow branch, i.e.
there is no shadow pole
connected with a deuteronlike state, in agreement with
the result of Ref.~\cite{Mor}. Instead,
above threshold there is a  second
crossing of $s$ with the linear part of the running mass (Fig.~2b). This
crossing is from below, which means that the
phase shift decreases slowly through 90$^\circ$. More generally,
the phase shift obtained from the  running mass
$m_d^2(s)$ is in accord with
Levinson's theorem (which holds for single
channel potential scattering), and
which requires the phase shift to drop by 180$^\circ$ from threshold to
$s=\infty$ to compensate for the existence of the deuteron pole.

{\bf 4. How to parametrize a resonance?}
As we have seen, for S-wave
resonances near the threshold resonance parameters depend on their
definition. In particular, the mass is quite different when defined
as the 90$^\circ$ BW mass or as the pole position.
Furthermore,  one may need two poles to describe the same resonance!
This, no doubt, is likely to cause
headaches for anyone involved with compilations such as Ref. \cite{PDG}.

The running mass $m^2(s)$ contains all
the essential information needed for the analytic
continuation to the poles. Together with the rather trivial imaginary
parts of eq.~(\ref{runmass}) it defines the propagator.
This then determines any of the other parameters one may wish to know:
the positions (and residues)
of the poles (one ore two), the coupling constant, the BW
mass, the BW width,  the scattering length and the
effective range parameter.

Near the threshold $m^2(s)$  depends on two parameters only, as
the examples in eqs.~(\ref{mBW}, \ref{deut}) demonstrate.
Eq.~(\ref{deut})  could be replaced by a form like
(\ref{mBW}) through a reparametrization of the T matrix element,
but then $m_{BW}^2-4m_K^2$ and $\gamma$ would have the
opposite sign and the physical  interpretation
would be lost. Therefore, for hadron-hadron states
one should use a form like
(\ref{deut}) (Fig.~2b) with its linear term having a slope $>1$.

Which two parameters should be
chosen to fix $m^2(s)$ is a question of taste,
provided they are not linearily
dependent\footnote{One pole and its residue is not sufficient
to fix the two parameters of the running mass,
because near the threshold
both of these depend to lowest order only on the quantity
$(m^2_{BW}-m_K^2)/\gamma $. Thus to fix the running mass one
must also give the second (shadow) pole or another parameter.},
but a natural choice would be:
(i) For  \qq\ state above threshold,
the conventional Breit Wigner $90^\circ$ mass and the (partial) width;
(ii) For  \qq\ state below threshold, the (second sheet) pole position
and partial width coupling parameter $\gamma $;
(iii) For a hadron-hadron state (below threshold),
the pole mass and the effective range $r$ parameter
(or the slope of $m^2(s)$ above threshold).
But, may I suggest that anyone making a fit to a resonance near
an  S-wave threshold should also compute the other relevant parameters
mentioned above.

In conclusion, for an S-wave
resonance I find the running mass to be the
natural quantity to be determined by experiment, which  should be the
best meeting ground for experiment, phenomenology and theory.
\eject

\eject
{\bf Figure captions.}
\bigskip

\noindent
Fig.~1

(a) The trajectories of the  poles in the complex plane of the c.m.
meson momentum $k_K$, when the bare meson mass is varied.
The dotted straight lines show the position of the bound state pole,
or the third sheet resonance pole and their shadow poles
 when $\Gamma_{\pi\eta}=0$. A filled square
is an example of the physical pole position,
while an unfilled square shows the
shadow pole. The points on the hyperbola show how the poles are shifted
when one adds a small finite $\Gamma_{\pi\eta}$ (15 MeV).
See text for  details.

(b) The same poles as in (a) but in the $s$-plane.
The full drawn curve shows
the physical pole positions and the dashed curve the shadow poles
when $\Gamma_{\pi\eta}=0$. The dotted curve show how
the poles are shifted when one adds a small
$\Gamma_{\pi\eta}$. Note that it  is the shadow pole
which turns into the "physical" third sheet resonance pole. The dotted
lower half parabola show the BW parameters ($m_{BW},-m\Gamma$).
See text for details.
\bigskip

\noindent
Fig.~2.

  (a) The running mass $m^2(s)$ of eq.~(\ref{mBW}) from which one can
read off the pole and shadow pole
positions for bound states and virtual bound
states of Fig. 2b, and in addition the BW  mass above threshold.
See text.

(b) The running mass for the deuteron fixed by the
experimental values for the scattering
length $a=3.82m_\pi^{-1}$ and effective range
$r=1.2m_\pi^{-1}$ parameters. The crossing point below
threshold gives the deuteron pole.
Note that the slope of the running mass above threshold
is $>1$ which is the signature for a
hadron-hadron bound state. See text.

\end{document}